\documentstyle[aps,
preprint, 
tighten,
multicol
]{revtex}
\begin{document}

\def\beqar{\begin{eqnarray}}
\def\eeqar{\end{eqnarray}}
\def\be{\begin{eqnarray}}
\def\ee{\end{eqnarray}}
\def\beqast{\begin{eqnarray*}}
\def\eeqast{\end{eqnarray*}}
\def\lag{\langle}
\def\rag{\rangle}
\def\fnote#1#2{\begingroup\def\thefootnote{#1}\footnote{#2}
\addtocounter{footnote}{-1}\endgroup}
\def\beq{\begin{equation}}
\def\eeq{\end{equation}}
\def\haf{\frac{1}{2}}
\def\pa{\partial}
\def\ca{{\cal A}}
\def\cb{{\cal B}}
\def\cc{{\cal C}}
\def\cd{{\cal D}}
\def\ce{{\cal E}}
\def\cf{{\cal F}}
\def\bB{\bar{B}}
\def\cH{{\cal H}_{\rm eff}}
\def\cN{{\cal N}}
\def\cR{{\cal R}}
\def\plb#1#2#3#4{#1, Phys. Lett. {\bf B#2}, #3 (#4)}
\def\npb#1#2#3#4{#1, Nucl. Phys. {\bf B#2}, #3 (#4)}
\def\prd#1#2#3#4{#1, Phys. Rev. {\bf D#2}, #3 (#4)}
\def\prl#1#2#3#4{#1, Phys. Rev. Lett. {\bf #2}, #3 (#4)}
\def\mpl#1#2#3#4{#1, Mod. Phys. Lett. {\bf A#2}, #3 (#4)}
\def\rep#1#2#3#4{#1, Phys. Rep. {\bf #2}, #3 (#4)}
\def\ll#1#2{\lambda_{#1}\lambda_{#2}}
\def\llp#1#2{\lambda_{#1}\lambda'_{#2}}
\def\lplp#1#2{\lambda'_{#1}\lambda'_{#2}}
\def\lplps#1#2{\lambda'_{#1}\lambda'^*_{#2}}
\def\lpplpp#1#2{\lambda''_{#1}\lambda''_{#2}}
\def\lpplpps#1#2{\lambda''_{#1}\lambda''^*_{#2}}
\def\llpp#1#2{\lambda_{#1}\lambda''_{#2}}
\def\lplpp#1#2{\lambda'_{#1}\lambda''_{#2}}
\def\elpp#1#2{\epsilon_{#1}\lambda''_{#2}}
\def\slash#1{#1\!\!\!\!\!/}
\def\rpv{\slash{R_p}~}
\def\rb{B\!\!\!\!/}
\def\rl{L\!\!\!/}
\def\ckm#1#2{V_{#1} V_{#2}^*}
\def\cur#1#2{\bar{#1} #2_{-}}


\draft
\preprint{
\begin{tabular}{r}
KIAS-P98039 \\
hep-ph/9811201
\end{tabular}
}
\title{
Implication of Super-Kamiokande Data on R-parity Violation 
}

\author{
Eung Jin Chun$\!\!$
\thanks{E-mail: ejchun@kias.re.kr}
and
Jae Sik Lee$\!\!$
\thanks{E-mail: jslee@kias.re.kr}
}

\address{
School of Physics, Korea Institute for Advanced Study, 
207-43 Cheongryangri-dong, Dongdaemun-gu, Seoul
130-012, Korea 
}

\maketitle

\begin{abstract}
R-parity violating  bilinear (soft) terms in the 
supersymmetric standard model would be the leading source for 
nonzero neutrino masses and mixing.
We point out that the mixing between neutralinos (charginos) and 
neutrinos (charged leptons) driven by the bilinear terms take
factorized forms, which may enable us to probe the neutrino 
mixing parameters in a collider.
It is then shown that the Super-Kamiokande data on atmospheric neutrinos
require all the baryon number violating couplings to be substantially 
suppressed: $\lambda''_{\rm any} <10^{-9}$.
\end{abstract}



The supersymmetric standard model  allows for
the baryon number B and the lepton number L violation. 
Such B and L violations are usually discarded by imposing 
the R-parity under which the superpotential is divided into
R-parity conserving and violating parts:
\beqar \label{Rviol}
&&h^u_i H_2 Q_i U^c_i + h^d_i H_1 Q_i D^c_i + h^e_i H_1 L_i E^c_i +
       \mu H_1 H_2;  \\
&&\lambda_{ijk}L_iL_jE_k^c+\lambda_{ijk}'L_iQ_jD_k^c+
\lambda_{ijk}''U_i^cD_j^cD_k^c    + \epsilon_i \mu L_i H_2  , \nonumber
\eeqar
where $i,j,k$ are generation indices.
There is, however, no compelling theoretical justification 
for this requirement.  Rather, explicit R-parity violation would well
be the origin of nonzero neutrino masses and mixing required by the 
recent observation of neutrino oscillation in 
the Super-Kamiokande \cite{SK-ATM}.
One of the essential features of L and $R_p$ violation 
is the generation of 
the mixing between neutralinos (charginos) and neutrinos (charged leptons).
These mixing come from the presence of $\Delta L=1$ 
bilinear soft terms which generate nonzero vacuum expectation values (VEVs)
for sneutrino fields unaligned with $\epsilon_i$ \cite{HSetc}.  
As we will see, the neutralino-neutrino and chargino-charged lepton mixing 
(denoted by $\Theta^N$ and $\Theta^{L,R}$, respectively) 
are factorized into $R_p$ conserving and violating parts 
as $\Theta_{ij}=c_i \xi_j$,
thereby leading to important phenomenological consequences. 
If the so-called tree level mass coming from the neutralino-neutrino 
mixing gives the dominant contribution, one can express the sizes of 
$\Theta$'s in terms of the atmospheric neutrino mass scale found in  
the Super-Kamiokande data.  Then, the factorization property 
enables us to relate the neutrino mixing parameters to the L violating 
sparticle decay rates to charged leptons, which are detectable in a collider.
It also turns out that the mixing $\Theta$ 
combined with the B violation can lead to a fast nucleon decay into 
a neutrino or charged lepton, unless all the B violating 
couplings $\lambda''$ are suppressed very strongly.

\medskip

Let us start our discussion on  the mixing $\Theta$ coming from the 
L violating bilinear terms.  The relevant soft supersymmetry (SUSY) 
breaking terms are
\beqar \label{Vsoft}
V_{\rm soft} &=& 
m^2_{H_1} |H_1|^2 +  m^2_{L_i} |L_i|^2  \nonumber  \\
   & +& (m^2_{L_iH_1} L_i H_1^\dagger + B H_1H_2 + B_i L_i H_2 + h.c.)  .
\eeqar
When the lepton number violating terms are small, 
$\epsilon_i, \lambda, \lambda' << 1$, as indicated by small neutrino 
masses advocated by current experiments, it is particularly convenient 
to use the linear approximation to rotate away 
the $\epsilon_i$ terms from the superpotential (\ref{Rviol}): 
\begin{eqnarray} \label{AWAY}
H_1 \rightarrow H_1'&=&H_1 + \epsilon_i L_i, \nonumber \\
L_i \rightarrow L_i'&=&L_i - \epsilon_i H_1  \,.
\end{eqnarray}
which is valid only up to ${\cal O}(\epsilon_i)$.  In this basis,
the lepton number is defined as in the supersymmetric limit, the merit of 
which is that the mixing $\Theta$ can be expressed only in terms of the basis
independent quantities $\langle \tilde{\nu_i}' \rangle = 
\langle \tilde{\nu_i} \rangle - \epsilon_i \langle H_1 \rangle$ 
as implied by Eq.~(\ref{AWAY}).
Upon the minimization of the scalar potential, 
the lepton number violating soft parameters $B_i', m^{'2}_{L_iH_2}$ induce 
nonzero VEVs $\langle\tilde{\nu}'_i\rangle \equiv -v_1 \xi_i$ \cite{CW}:  
\beq \label{Ai}
\xi_i\equiv\frac{B_i'\tan\beta+{m}^{'2}_{L_iH_1}}
              {m_{L_i}^2+M_Z^2\cos{2\beta}/2 }\,,
\eeq
where $\tan\beta=v_2/v_1$, $v_i =\langle H_i \rangle$ and $v=174$ GeV.  
Due to the sneutrino VEVs, neutrinos mix with neutralinos in the $7\times
7$ mass matrix in the $(\nu_i; \tilde{B}, \tilde{W}_3, \tilde{H}_1^0, 
\tilde{H}_2^0)$ basis:
\beq \label{7x7}
 \pmatrix{ m_\nu^{\rm loop} & m_D \cr  m_D^T & M_N \cr} 
\eeq
where $m_{\nu}^{\rm loop}$ comes from 1-loop diagrams involving the trilinear 
couplings $\lambda',\lambda$, and $M_N$ is the usual neutralino mass matrix. 
Since the $3\times4$ matrix $m_D$, 
\beq
m_D=\pmatrix{M_Z s_W \xi_1 \cos\beta & -M_Z c_W \xi_1 \cos\beta & 0 &0 \cr
             M_Z s_W \xi_2 \cos\beta & -M_Z c_W \xi_2 \cos\beta & 0 &0 \cr
             M_Z s_W \xi_3 \cos\beta & -M_Z c_W \xi_3 \cos\beta & 0 &0 \cr},
\eeq
is much smaller than $M_N$, it is enough to use see-saw formula 
to find the mixing matrix
$\Theta^N$ between weak eigenstate fields 
$( \tilde{B}, \tilde{W}_3, \tilde{H}_1^0, \tilde{H}_2^0)$ and $\nu_j$: 
\beqar \label{TheN}
&&\Theta^N_{ij}=-(M_N^{-1}m_D^T)_{ij} \equiv {M_Z \over F_N} b_i 
        \xi_j \cos\beta \quad\mbox{with} \nonumber \\
&&b_1 = -{s_W  M_2 \over M_1 c_W^2+ M_2 s_W^2}, \quad
b_2= {c_W  M_1 \over M_1 c_W^2+ M_2 s_W^2}, \nonumber\\
&&b_3 =  -\sin\beta {M_Z\over \mu }, \quad\qquad~~
b_4= \cos\beta {M_Z\over \mu },  \nonumber \\
&&F_N = {M_1 M_2 \over M_1 c_W^2+M_2 s_W^2}
       +{M_Z^2\over \mu} \sin2\beta. 
\eeqar
Here $s_W=\sin\theta_W$, etc, and $\theta_W$ is the weak mixing angle. 

In a similar way, the mixing between charged leptons and
charginos can be obtained from the following $5\times5$ mass matrix
for the fields, $(l_i^{\pm}; \tilde{W}^{\pm}, \tilde{H}^{\pm}_{2,1})$:
\beqar \label{5x5}
\pmatrix{m_C & m_L \cr m_R & M_C \cr}~\mbox{with}~
m_L&=&\pmatrix{-\sqrt{2}M_W \xi_1 \cos\beta & 0 \cr 
               -\sqrt{2}M_W \xi_2\cos\beta & 0 \cr
               -\sqrt{2}M_W \xi_3\cos\beta & 0\cr },  \\
m_R&=&\pmatrix{ 0 & 0 & 0 \cr m_e \xi_1 &  
 m_\mu \xi_2 & m_\tau \xi_3\cr }, \nonumber
\eeqar
where $m_C$ is the diagonal charged lepton mass matrix 
with elements $m_{l_i}$,
and $M_C$ is the usual chargino matrix.  
While the charged lepton masses $m_C$ remain untouched 
to a good approximation, 
the left-handed charginos ($\tilde{W}^-,\tilde{H_1}^-$) mix with
the left-handed leptons ($e,\mu,\tau$) through the matrix,
\beqar  \label{TheL}
\Theta^L_{ij} &=& -(m_C M_C^{-1})^T_{ij} \equiv  c^L_i \xi_j \cos\beta, 
        \nonumber\\
\quad\mbox{with}\quad\nonumber 
c^L_1&=&\sqrt{2}{M_W\over F_C},\quad 
c^L_2=-2\sin\beta{M_W^2\over\mu F_C}, \\
F_C &=& M_2+M_W^2\sin2\beta/\mu .
\eeqar
The mixing between the charginos ($\tilde{W}^+,\tilde{H_2}^+$) and the
anti-leptons ($e^c,\mu^c,\tau^c$) is given by the matrix,
\beqar  \label{TheR}
\Theta^R_{ij} &=& \left(M_C^{-1} (\Theta^L m_C-m_R) \right)_{ij}
\equiv c^R_i {m_{l_j} \xi_j \cos\beta\over F_C},\nonumber\\
\quad\mbox{with}\quad
c^R_1&=&-\sqrt{2}(1-\cot\beta{\mu\over M_2}){M_W M_2\over\mu F_C}, \nonumber \\
c^R_2&=&{2\over\cos\beta}(1+\cos^2\beta{M_W^2\over M^2_2}){M_2^2\over\mu F_C}
.
\eeqar
Note that $\Theta^R$ contains the lepton mass suppression.

The L violating processes arising from the mixing
(\ref{TheN},\ref{TheL},\ref{TheR}) 
show also the factorization property, which
may provide a striking connection to the neutrino physics, as we will see.
Here, we should mention that the physical quantities
involving the mixing of $\tilde{W}$ or $\tilde{H}_1$ are not given by
$\Theta^N$ or $\Theta^L$ alone but by the  $\Theta^N_{3j}-\Theta^L_{2j}$. 
Recall then that in the basis where $\epsilon_i$
are not rotated away \cite{NP}, $\Theta^N_{3j}$ and $\Theta^L_{2j}$ 
contain extra $\epsilon_j$ which are canceled out in the above combination.  
Therefore, in either basis, the quantities $\xi_i$ dictate the properties of 
the neutralino-neutrino and chargino-charged 
lepton mixing, and thus the corresponding neutrino masses shown below.

\medskip

The see-saw reduced mass matrix from (\ref{7x7}) gives so-called the
tree level neutrino masses:
\beq \label{TREE}
m_{\nu_{ij}}^{\rm tree}
=\frac{M_Z^2 }{F_N} \xi_i \xi_j \cos^2\beta,
\eeq
which, as is well-known, gives a mass to only one neutrino.
If the tree mass (\ref{TREE}) dominates over the loop mass 
$m_{\nu}^{\rm loop}$,  the heaviest neutrino mass is given by
$m_{\nu_3} \equiv M_Z^2 \xi^2 \cos^2\beta/F_N$ with 
$\xi=\sqrt{\xi_1^2+\xi_2^2+\xi_3^2}$. 
Now the matrix $\Theta^N$ can be written as
\beq
\Theta^N_{ij}={b_i \xi_j\over \xi}\left(m_{\nu_3}\over F_N\right)^{1/2} ,
\eeq
which shows that neutralinos mix only with the heaviest neutrino $\nu_3$.
Here we take $m_{\nu_3}\simeq \sqrt{\Delta m^2_{\rm atm}} 
\simeq 5\times 10^{-2}$ eV as implied by the Super-Kamiokande data 
\cite{SK-ATM}, and then one finds 
\beq
\xi \cos\beta \simeq 0.7\times10^{-6} \left(F_N\over M_Z\right)^{1/2} 
          \left(m_{\nu_3} \over 0.05 {\rm eV}\right)^{1/2} .
\eeq
In addition, the mass matrix (\ref{TREE}) determines two mixing angles 
related to the heaviest neutrino \cite{JN} and thus gives rise to
the oscillation amplitudes with $\Delta m^2_{\rm atm}=\Delta m^2_{32,31}$  
as follows:
\beqar \label{AMP}
\sin^2\theta^{\rm atm}_{ee\!\!/} &=& 
       4{\xi_1^2\over \xi^2}(1-{\xi^2_1\over \xi^2}), \nonumber\\
\sin^2\theta^{\rm atm}_{\mu\tau} &=& 
       4{\xi_2^2\over \xi^2}{\xi_3^2\over \xi^2} .
\eeqar
The other mixing angles related to the solar neutrino oscillation can be 
found only after taking the next leading masses $m_{\nu}^{\rm loop}$.
The Super-Kamiokande data \cite{SK-ATM} requires large mixing for 
$\nu_\mu-\nu_\tau$ oscillation, and thus $\xi \sim \xi_2 \sim \xi_3$.
A limit on $\xi_1^2/\xi^2$ can come from the CHOOZ $\bar{\nu}_e$ disappearance 
experiment \cite{CHOOZ}: $\xi_1^2/\xi^2 \lesssim 0.05$ \cite{BPWW}
which is applicable for $\Delta m^2_{\rm atm} \gtrsim 
2\times 10^{-3} {\rm eV}^2$. 
It is now important to notice that $\xi_i$'s appear also in the 
mixing $\Theta^{L,R}$, which can lead to  L violating sparticle decays.  
Of particular interest is the lightest sparticle (LSP) decay with same-sign 
dilepton signal ($l_i l_j$ (+ 4 jets)) inside a collider \cite{SIGNAL}.  
Typical example is the two body decay, $\chi_1^0 \to l_j W$ \cite{MRV}. 
Having the decay rate,
\begin{equation} 
\Gamma(l_j) \sim G_FM_{\chi_1^0}^3 |\xi_j|^2 \cos^2\beta/2\pi, 
\end{equation}
its decay length is roughly $\lesssim 1$ mm.  Then, 
one will be able to probe the neutrino oscillation amplitudes (\ref{AMP})
by counting the same-sign lepton events. 
For example, in the $e^+e^-$ collider, taking $\tau$ in one hemisphere in the
direction of incident $e^-$ and counting the same-sign leptons in the
other hemisphere gives the relation
\beq
(e\tau):(\mu\tau):(\tau\tau) =\xi_1^2:\xi_2^2:\xi_3^2.
\eeq
The CHOOZ result implies that 
${(e\tau)\over(\mu\tau)+(\tau\tau)} \lesssim 0.05$ 
independently of $\Delta m^2_{\rm atm}$.
Otherwise, one can conclude that 
$\Delta m^2_{\rm atm} < 2\times10^{-3} {\rm eV}^2$.
For $M_{\chi^0_1} < M_W$,  the LSP has only three body
decay modes like $\chi^0_1 \to l_j+2 {\rm jets}$ through $W/Z$ boson 
exchanges which can barely occur inside a collider.
To get a qualitative estimation of the decay length, 
let us take bino $\tilde{B}$ as the LSP.  
Its decay rate normalized by the muon decay
rate is then given by $\Gamma_{\chi^0_1}/\Gamma_\mu \approx  13.4
b_1^2 (m_{\nu_3}/F_N) (M_{\chi^0_1}/m_\mu)^5$ where the numerical factor
comes from the final state summation.  If one assumes the unification
relation $M_1={5\over3} t_W^2 M_2$, one gets $b_1 = 3/8s_W$,
$F_N =  3M_1/8 s_W^2$, and thus 
\begin{equation}
 \tau_{\chi^0_1} \sim 0.7 m 
 \left(0.05~{\rm eV} \over m_{\nu_3}\right)
 \left(M_W \over M_{\chi_1^0}\right)^4. 
\end{equation}
Of course, details will depend 
on the mass parameters in the neutralino sector \cite{ING}.

\medskip

Dominance of the tree mass over the loop mass is generally true if 
there is no alignment in soft masses  \cite{BGNN}.  But, 
the loop-to-tree mass ratio is then too small for the loop mass to 
account for the solar neutrino problem. 
The situation can be different if supersymmetry breaking process is 
flavor blind as in the framework of minimal supergravity,
or gauge mediated supersymmetry breaking, which is 
usually required for suppressing flavor changing processes 
in supersymmetric theories.  In this framework,   the parameters
$B'_i$ and ${m'}^2_{L_i H_1}$ are zero at the mediation scale of supersymmetry
breaking $M_m$, but their nonzero values (which are linearly dependent on 
$\lambda,\lambda'$ or $\epsilon_i$ given at $M_m$) are generated at the 
weak scale through renormalization group (RG) evolution \cite{NAR}, 
and thus both the tree and loop mass are of the radiative origin.  
In order to get a rough idea how nonzero values of $\xi_i$ are generated 
and are related to the R-parity violating couplings in Eq.~(\ref{Rviol}), 
let us integrate the relevant RG equations \cite{CHUN2} in a crude 
way to get, {\it e.g.},  
\beq \label{pisi}
 \xi_i  \sim 
 {3 \lambda'_{i33}h_b \over 8\pi^2} \ln{M_m \over M_Z}
\eeq
where $h_b$ is the b-quark Yukawa coupling.  In our basis, rotating away
$\epsilon_i$ terms induces additional  contributions to trilinear couplings,
in other words, e.g., $\epsilon_i h_b$ has to be added to $\lambda'_{i33}$.
{}From Eq.~(\ref{pisi}),
one can see that the smallness of $\xi_i \sim 10^{-6}$ is due to 
both the loop factor involving the Yukawa coupling $h_b$, and small input 
couplings $\lambda',\lambda$ or $\epsilon_i$
which would originate from a horizontal symmetry \cite{Hori,Hwang}.
As the tree mass contains the logarithmic enhancement factor, it
is typically much larger than the loop mass.
It has been recently found that the loop mass can also be enhanced for large 
$\tan\beta$ to explain the solar neutrino mass scale \cite{Hwang,JV,KN}.
On the other hand, there could be a cancellation in $\xi_i$ (\ref{Ai}) to 
suppress the tree mass substantially \cite{CHUN2}.
In this case, the tree mass could be comparable to the loop mass,  and thus
our formulae (\ref{AMP}) are invalidated and become complicated functions
of the input parameters $\lambda, \lambda'$ or $\epsilon_i$.

\bigskip

Another immediate consequence of the mixing $\Theta$ is 
proton and neutron decay when combined with the B violating couplings 
$\lambda''$. The bounds from proton longevity are well 
studied when both $\lambda'$ and $\lambda''$ are present.
The strongest bounds come from tree level diagrams with exchanges of 
a right-handed squark $\tilde{d}^k_R$ \cite{HK}
\beq \label{old1}
 \lambda'_{ijk} \lambda''_{11k} \lesssim 10^{-24} \qquad j=1,2 \,,
\eeq
for a squark mass of 1 TeV.  The diagrams for the proton decay at one loop
level involve the other couplings and give rise to mild constraints \cite{SV}
\beq \label{old2}
 \lambda' \lambda'' \lesssim 10^{-9} \,.
\eeq
New bounds on $\lambda''$ combined with the R-parity violating bilinear terms
were first derived in Ref.~\cite{BP} by using a mass insertion due to
$L_i H_2$ terms in the superpotential, which yields
\beq \label{old3}
  \epsilon_i  \lambda''_{112} \lesssim 10^{-21} \,.
\eeq
To take into account the effect of the R-parity violating bilinear terms
consistently, one has to consider the full mixing
between leptons and neutralinos or charginos discussed above, and
derive the bounds on the combination $\lambda'' \Theta_i$.
It turns out then that there are some more new diagrams at tree level 
contributing to the the proton or neutron decay  and provide more stringent
bounds.  

To derive the four-fermion effective Lagrangian responsible for the proton 
decay to a neutrino resulting  from the exchange of right-handed squarks,
let us first write down explicitly the interaction Lagrangian between 
neutralino, quark and right-handed squark
\beqar
&& \left\{ \bar{u}\left[ \sqrt{2} g Q_u\tan\theta_W P_L\tilde{B}-
h^u_1 P_R\tilde{H}_2^0\right]\tilde{u}_R\right.
\nonumber \\
&&+\left.
\bar{d}\left[ \sqrt{2} g Q_d\tan\theta_WP_L\tilde{B} -
h^d_1 P_R\tilde{H}_1^0\right]\tilde{d}_R\right\} +\rm{h.c.},
\eeqar
where $P_{L,R}=(1\mp\gamma_5)/2$, and  $h^u_i=g m^u_i/\sqrt{2}M_W \sin\beta$,
$h^d_i=g m^d_i/\sqrt{2}M_W \cos\beta$ are the up and down type quark Yukawa
couplings respectively.
The R-parity violating  interactions involving $\lambda''$ are
\beq \label{lam''}
\lambda''_{ijk}\left[(\overline{u_R^i})^cd_R^j\tilde{d}_R^k+
(\overline{u_R^i})^cd_R^k\tilde{d}_R^j+
(\overline{d_R^j})^cd_R^k\tilde{u}_R^i \right].
\eeq
{}From the above two equations,  one obtains 
the four-fermion interactions describing the proton decay
to a neutrino,
\beqar \label{ff1}
&&\frac{\sqrt{2}g\lambda''_{112}}{m_{\tilde{u}_R}^2}
\overline{s^c}P_Rd\left[(2Q_u t_W)\overline{\tilde{B}}P_R u-
\frac{m_u}{M_W\sin\beta}\overline{\tilde{H}_2^0}P_Lu\right] 
\nonumber \\
&+&\frac{\sqrt{2}g\lambda''_{112}}{m_{\tilde{d}_R}^2}
\overline{u^c}P_Rs\left[(2Q_dt_W)\overline{\tilde{B}}P_R d-
\frac{m_d}{M_W\cos\beta}\overline{\tilde{H}_1^0}P_Ld\right] \\
&+&\frac{\sqrt{2}g\lambda''_{112}}{m_{\tilde{s}_R}^2} \nonumber
\overline{u^c}P_Rd\left[(2Q_dt_W)\overline{\tilde{B}}P_R s-
\frac{m_s}{M_W\cos\beta}\overline{\tilde{H}_1^0}P_Ls\right],
\eeqar
where $m_{\tilde{q}}$ are the masses of the exchanged squarks.
Eq.~(\ref{ff1}) leads to the hadronic matrix elements for proton and kaon
described by the following operators:
\beqar \label{chiop}
&&\ca~\Theta^N_{1j}(\overline{\nu_j}P_R P) K^- +
\cb~\Theta^N_{3j}(\overline{\nu_j^c}P_L P) K^- +
\cc~\Theta^N_{4j}(\overline{\nu_j^c}P_L P) K^-  \nonumber \\
&& \mbox{where}~~
\ca=\frac{4\sqrt{2}g\lambda''_{112}\eta}{3\tilde{m}^2}
\frac{\Delta\tilde{m}^2}{\tilde{m}^2},
~\cb=\frac{\sqrt{2}g\lambda''_{112}\eta}{\tilde{m}_d^2}
         \frac{m_d+m_s}{M_W\cos\beta}, \nonumber\\
&&~~~~~~~~~\cc=\frac{\sqrt{2}g\lambda''_{112}\eta}{\tilde{m}_d^2}
\frac{m_u}{M_W\sin\beta}.  
\eeqar
Here $\tilde{m}^4=m^2_{\tilde{u}_R}m^2_{\tilde{d}_R}$ ,
$\Delta\tilde{m}^2=m^2_{\tilde{u}_R}-m^2_{\tilde{d}_R}$
and we took $m^2_{\tilde{d}_R}=m^2_{\tilde{s}_R}=\tilde{m}_d^2$.
Using the naive dimensional analysis rule of Ref.~\cite{NDA}, we estimate the
size of the hadronic coefficient $\eta$ as $|\eta|\simeq 4\pi f_{\pi}^2$ with
$f_{\pi}=93$ MeV. Note that contributions from $\cb$ and $\cc$ to the proton
decay would be suppressed comparing with those from $\ca$ because of the quark
mass suppressions in $\cb$ and $\cc$ 
unless $\Delta\tilde{m}^2$ is very small and $\tan\beta$ is very large. 
Let us make some more remarks on the terms in Eq.~(\ref{chiop}).
The $\cb$ term is nothing but the term arising from the trilinear
coupling $\lambda'_{j11}$ or $\lambda'_{j22}$ defined in the mass
basis of the fields.  In other words, $\lambda'_{j11}$ is the induced coupling
of the size $\epsilon_j h^d_1$ or $\Theta^N_{3j} h^d_1$.  On the other hand,
the $\ca$ and $\cc$ terms cannot be related to trilinear couplings 
$\lambda'$ and thus provide the genuine new contributions of the bilinear
R-parity violation to proton decay. 
In a naive mass insertion method worked out in Ref.~\cite{BP}, only the $\cc$ 
term can be obtained.

{}From the mixing (\ref{TheN}) of the neutralinos 
with  the neutrinos, we obtain from Eq.~(25) 
the decay rate of the proton into 
$K^+$ and $\nu_3$ neglecting $\cb$ and $\cc$ as follows
\beq
\Gamma=\frac{|\ca|^2}{32\pi}
\frac{(m_P^2-m_K^2)^2}{m_P^3}{b_1^2 m_{\nu}\over F_N},
\eeq
where $m_P$ denotes the proton mass, {\rm etc}.
{}From the unobservation of the decay $P\rightarrow K\nu$ \cite{pdg}, 
we obtain a very stringent bound,
\beqar \label{ONE}
&&\lambda''_{112} < 6 \times 10^{-19} ~x \quad{\rm where}\quad
x=\left(\frac{0.05~{\rm eV}}{m_{\nu}}\right)^{1/2} \\ && \nonumber
\left(\frac{F_N}{300~{\rm GeV}}\right)^{1/2}
\left(\frac{\tilde{m}}{300~{\rm GeV}}\right)^2
\left(\frac{\tilde{m}^2}{b_1\Delta\tilde{m}^2}\right).
\eeqar
Note that this bound corresponds to that with $\lambda'\sim 10^{-7}$
in Eq.~(\ref{old1}) or $\epsilon_i \sim 10^{-4}$ in Eq.~(\ref{old3}).

\bigskip

We can also find that the chargino--charged lepton mixing $\Theta^{L,R}$ 
allows for more {\it tree} level decay of neutron or proton to charged leptons 
involving trilinear couplings other than $\lambda''_{112}$. 
This can be seen by combining the $\lambda''$ couplings (\ref{lam''}) and
the interaction of chargino, quark and 
right-handed squark described by
\beq
V_{ij}^* h_j^{d} \overline{\tilde{H}_1^+} P_L u^i \tilde{d}_R^{j*}
+V_{ij} h_i^{u} \overline{\tilde{H}_2^-} P_L d^j \tilde{u}_R^{i*}
+{\rm h.c.},
\eeq
where $V_{ij}$ are the CKM matrix elements.
The relevant four-fermion operator involving charginos takes
the form;
\beq
\frac{\lambda''_{ijk} V_{ip} h_i^{u} }{m_{\tilde{u}_R^i}^2}
\overline{d_j^c} P_R d_k \overline{\tilde{H}_2^-} P_L d_p
+\frac{\lambda''_{ijk} V_{kp}^* h_k^{d} }{m_{\tilde{d}_R^k}^2}
\overline{u_i^c} P_R d_j \overline{\tilde{H}_1^+} P_L d_p \,,
\eeq
from which one finds 
the corresponding effective couplings for nucleons and mesons;
\beqar \label{npdecay}
&&\cd~\Theta^R_{2j}(\overline{l_j}P_L N) K^- +
\ce~\Theta^L_{1j}(\overline{l_j^c}P_L P) \pi^0+
\cf~\Theta^L_{1j}(\overline{l_j^c}P_L P) K^0 , \nonumber \\
&&\mbox{where} \quad 
\cd=\frac{\lambda''_{212} V_{21} h_2^{u}\eta}{m_{\tilde{u}_R^2}^2}+
\frac{\lambda''_{312} V_{31} h_3^{u}\eta}{m_{\tilde{u}_R^3}^2}, \\
&&~~~~~~~~~~
\ce=\frac{\lambda''_{113} V_{31} h_3^{d}\eta}{m_{\tilde{d}_R^3}^2},~
\cf=\frac{\lambda''_{123} V_{31} h_3^{d}\eta}{m_{\tilde{d}_R^3}^2}. \nonumber
\eeqar
As in the previous case, the $\ce$ or $\cf$ term is again related to the 
$\lambda'$ coupling of the size $\epsilon_j V_{31} h^d_3$ or 
$\Theta^L_{1j} V_{31} h^d_3$.  Note however that the $\cd$ term gives rise to
a new tree level contribution involving $\lambda''_{k12}$.
Unobservation of the above decays \cite{pdg} leads to the very
severe constraints on $\lambda''_{212}$, $\lambda''_{312}$,
$\lambda''_{113}$, and $\lambda''_{123}$ as given in Table \ref{haha1}.
There, the normalization factors $w,y,z$ are given by
\beqar \label{FAC}
w&=&{\tilde{m}^2_{b_{_R}} \mu F_C \over (m_{\nu}{F_N})^{1/2}} 
    \left(\frac{\xi}{\xi_2}\right) \cot\beta \nonumber  \\
y,z&=&{\tilde{m}^2_{c_{_R},t_{_R}} \mu  \over (m_{\nu}{F_N})^{1/2}} 
      { F_C^2 \over M_2^2}
    \left(\frac{\xi}{\xi_2}\right)\sin2\beta 
\eeqar
where $m_{\nu_3}$ is normalized by $5\times10^{-2}$ eV and the other mass 
parameters by 300 GeV as in Eq.~(\ref{ONE}). 
We considered the nucleon decay into muon for which $\xi/\xi_2 \sim 1$.
It would be worth emphasizing  again that one obtains very strong bounds on 
$\lambda''_{212}, \lambda''_{312}$; 
\beqar
   \lambda''_{212} &<& 2 \times 10^{-13} y  \nonumber\\ 
   \lambda''_{312} &<& 3 \times 10^{-14} z  \,. 
\eeqar
which arise from the charged lepton and chargino mixing.

\bigskip 


Now taking into account the one loop diagrams with radiative corrections 
to the $\lambda''$ vertex, we can constrain the remaining $\lambda''$'s
\cite{SV}.
Relative to the tree diagram, one loop diagrams involving
$\lambda''_{ijk}$ will be suppressed by the factor $\zeta_{ijk}$, more
explicitly, $ (A_{\rm loop}^{ijk}/\lambda''_{ijk})=\zeta_{ijk}
(A_{\rm tree}/\lambda''_{112}) $ 
where $A_{\rm loop}$ and $A_{\rm tree}$ stand for the loop and tree
amplitudes, respectively. The one loop diagrams with the charged higgs boson
give
\beqar
\zeta_{ij1}\approx\frac{g^2m^u_im^d_j}
{16\pi^2M_W^2\sin 2\beta}V_{i2}^*V_{1j},
\nonumber \\
\zeta_{ij2}\approx\frac{g^2m^u_im^d_j}
{16\pi^2M_W^2\sin 2\beta}V_{i1}^*V_{1j}.
\eeqar
We derive the above suppression
factors using the following up-type quark, down-type quark
and charged higgs interaction
Lagrangian :
\beq
V_{ij}H^+\left[h_i^{u}\cot\beta\bar{u}_i P_L d_j
+h_j^{d}\tan\beta \bar{u}_i P_R d_j \right]  +{\rm h.c.}.
\eeq
With this one loop suppression factors the upper bounds on $\lambda''_{ijk}$
are given by
\beqar
\lambda''_{ijk} &<& \frac{6 \times 10^{-19}}{\zeta_{ijk}}
\left(\frac{5\times 10^{-2}~{\rm eV}}{m_{\nu}}\right)^{1/2} \\
&&\left(\frac{F_N}{300~{\rm GeV}}\right)^{1/2}
\left(\frac{\tilde{m}}{300~{\rm GeV}}\right)^2              \nonumber
\left(\frac{\tilde{m}^2}{b_1\Delta\tilde{m}^2}\right) \sin2\beta \,.
\eeqar
{}From these one-loop suppression factors,
we obtain the upper bounds on all $\lambda''_{ijk}$'s listed
in Table \ref{haha1}. The normalization factor $x'$ 
is given by $x'=x \sin2\beta$.  Note that the bounds from 1-loop diagrams
with the neutrino (charged lepton)--neutralino (chargino) mixing
are more stringent than any other existing bounds;
$\lambda'\lambda'' <10^{-9}$ \cite{SV}, 
$\lambda_{i33}\lambda''_{\rm all} <10^{-11}$ \cite{BP,SV}, or the
single bounds $\lambda''_{\rm all} <10^{-7}$ in  gauge mediated SUSY breaking
models \cite{CCL}.


\bigskip

In conclusion, we investigated the consequences of R-parity violation
when the bilinear soft breaking terms  are the leading source for 
the neutrino masses and mixing implied by the recent Super-Kamiokande 
data on atmospheric neutrinos.  Non-zero sneutrino VEVs lead to the mixing
between neutralinos (charginos) and neutrinos (charged leptons) which
have the factorization property. We show that, due to this property, 
neutrino mixing angles can be directly measured in colliders 
by comparing the branching ratios of the LSP.  
Taking into account the particle-sparticle mixing properly, 
we are able to find some more new contributions to proton and neutron decay
processes which yield very strong upper bounds on the B violating couplings.


\begin{table}
\caption{\label{haha1}
Constraints on $\lambda''_{ijk}$ 
obtained by combining the proton decay limits and the Super-Kamiokande
neutrino data. The normalization factors $x,w,y,z,x'$ can be found 
in the text.
}
\vspace{0.1 cm}
\begin{tabular}{cl}
Coupling  & Constraint\\
\hline 
$\lambda''_{112}$ & $6\times~10^{-19}$ $x$\\
$\lambda''_{113}$ & $3\times~10^{-15}$ $w$\\
$\lambda''_{123}$ & $7\times~10^{-15}$ $w$\\
$\lambda''_{212}$ & $2\times~10^{-13}$ $y$ \\
$\lambda''_{312}$ & $3\times~10^{-14}$ $z$\\
$\lambda''_{213}$ & $1\times~10^{-10}$ $x'$\\
$\lambda''_{223}$ & $5\times~10^{-10}$ $x'$\\
$\lambda''_{313}$ & $2\times~10^{-11}$ $x'$\\
$\lambda''_{323}$ & $1\times~10^{-10}$ $x'$\\
\end{tabular}
\end{table}

\end{document}